\documentclass[prd,twocolumn,showpacs,preprintnumbers,amsmath,amssymb]{revtex4}
\usepackage{graphicx}
\usepackage{epsfig}
\usepackage{rotating}
\usepackage{dcolumn}
\begin{document}
\preprint{IC/2002/81} 
\preprint{IISc/CTS/10-02}   
\preprint{CERN-TH/2002-288} 
\preprint{hep-ph/0211136}
\title{\bf Study of CP Property of the Higgs at a Photon Collider 
using $\gamma\gamma\rightarrow t\bar t\rightarrow l X $}
\author{ Rohini M. Godbole\footnote{On leave of absence from the Center for
Theoretical Studies, IISc, Bangalore, India.}}
\email{rohini.godbole@cern.ch}
\affiliation{CERN, Theory Division, CH-1211, Geneva 23, Switzerland}
\author{Saurabh D. Rindani\footnote{Permanent address : Physical Research
 Laboratory, Navrangpura, Ahmedabad 380 009, India}}
\email{saurabh@prl.ernet.in}
\affiliation{Abdus Salam International Centre for Theoretical Physics, Strada
Costiera 11, 34014 Trieste, Italy}
\author{Ritesh K. Singh}
\email{ritesh@cts.iisc.ernet.in}
\affiliation{Center for Theoretical Studies, IISc, Bangalore, India}
\begin{abstract}
We study possible effects of CP violation in the Higgs sector on $t\bar t$ 
production at a $\gamma\gamma$-collider.  These studies are performed in a  
model-independent way in terms of six form-factors 
$\{\Re(S_{\gamma}), \Im(S_{\gamma}), \Re(P_{\gamma}), 
\Im(P_{\gamma}), S_t, P_t\}$  which parametrize the CP mixing 
in the Higgs sector, and a strategy for their determination is developed.
We observe that the angular distribution of the decay lepton from $t/\bar t$ 
produced in this process is independent of any CP violation  
in the $tbW$ vertex and hence best suited for studying CP mixing in the  
Higgs sector. Analytical expressions are obtained for the angular distribution 
of leptons in the c.m. frame of the two colliding photons for a  general 
polarization state of the incoming photons. We construct combined asymmetries
in the initial state lepton (photon) polarization and the final state lepton 
charge.  They  involve CP even ($x$'s) and odd ($y$'s) combinations of the 
mixing parameters. We study limits up to which the values of $x$ and $y$, 
with only two of them allowed to vary at a time, can be probed by 
measurements of these
asymmetries, using circularly polarized photons. We use the numerical
values of the asymmetries predicted by various models to discriminate among
them. We show that this method can be 
sensitive to the loop-induced CP violation in the Higgs sector in the MSSM.
\end{abstract}
\maketitle
\noindent
\section{Introduction}
\begin{figure}
\begin{center}
\includegraphics[scale=0.60]{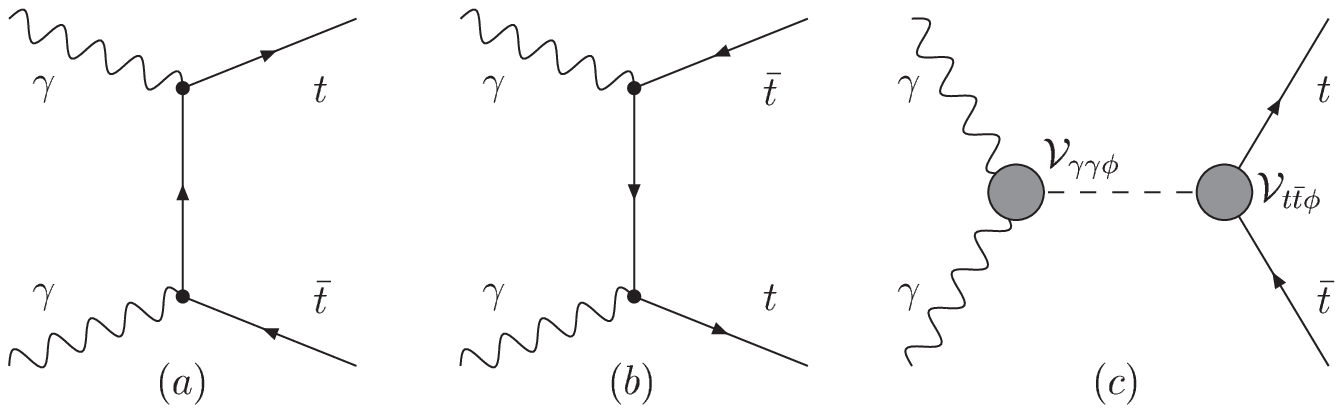}
\caption{\label{feynman}
Feynman diagrams contributing to $\gamma \gamma \rightarrow t \bar t$ production.}\end{center}
\end{figure}

The Standard Model (SM) has been tested to an extremely high degree of 
accuracy, reaching its high point in the precision measurements at LEP. 
However, the bosonic sector of the SM in not yet complete, the Higgs boson is 
yet to be found. A direct experimental demonstration of the Higgs mechanism 
of the fermion mass generation still does not exist. Also lacking is a first 
principle understanding of CP violation in the SM. In this note we look at 
the possibilities of probing  potential CP violation in the Higgs sector at 
the proposed $\gamma \gamma $ colliders \cite{tdr}. \\\\
Such a study necessarily means that we are looking at models
with an extended Higgs sector. CP violation in the Higgs sector can be either
explicit, one of the first formulations of such a CP violation 
being the Weinberg Model \cite{weinberg},  or can be 
spontaneous \cite{tdlee}, where the vacuum becomes CP non-invariant.
The mechanism for creating CP violation in the Higgs sector 
could be different in different models but all such mechanisms will result in 
CP mixing and then the mass eigenstate scalar will have no definite 
CP transformation property.  In specific models with an 
extended Higgs sector, such as the Minimal Supersymmetric Standard 
Model (MSSM), for example, the lightest Higgs remains more or less a CP 
eigenstate  and the two heavier states $H,A$, which would be CP-even and CP-odd
respectively in the absence of CP violation  and are close in mass
to each other, mix.  The expected mixing can actually be calculated as a 
function of the various parameters of the model~\cite{mssmpaper,asak1,newmssm}. 
In our study, however, we do not stick to a particular model of CP violation 
and adopt a model-independent approach to study the effects of this CP
violation on  $t \bar t$ production in $\gamma \gamma$ 
collisions. Such an approach has been adopted in earlier 
studies \cite{asakawa}. \\\\ 
We study $\gamma \gamma \rightarrow t \bar t$ through the diagrams shown in  
Fig. \ref{feynman}. It has been observed earlier \cite{asakawa}
that there exists a polarization asymmetry of the $t \bar t$ produced in the 
final state if the scalar $\phi$  exchanged in the $s$-channel is not a CP 
eigenstate. We parametrize the ${\cal V}_{\gamma \gamma \phi}, 
{\cal V}_{t \bar t \phi}$ vertices in a model-independent way in terms of 
six form factors to include  the CP mixing, following Ref.~\cite{asakawa}.  
We investigate in this study the  effect of such a CP mixing on 
the angular and charge asymmetries for the decay leptons coming from the 
$t/\bar t$ which reflect the top polarization asymmetries. 
It is known \cite{gunion} that $\gamma \gamma$ colliders 
can provide crucial information on the CP property of the scalar produced in
the $s$-channel, due to the very striking  dependence of the process on the
polarization of the $\gamma$'s.  These colliders will also offer the 
possibility of measuring the two-photon width of the SM Higgs very 
accurately \cite{sm2gh,maria}. The $\gamma \gamma$ production of a scalar 
followed by its decay into a $Z$ pair is shown to provide crucial information
required for a model independent confirmation of its spin and 
parity~\cite{zerwasnew}.  Possibilities of studying the MSSM
Higgs bosons in $\gamma \gamma$ collisions in the $b \bar b$ and 
neutralino-pair final states, are shown \cite{zerwas} to give access to regions 
of the supersymmetric parameter space not accessible at other colliders.
Thus in general the $\gamma \gamma$ 
colliders will provide a very good laboratory for studying the scalar sector. 
Here we concentrate on the polarization asymmetries of the
final state $t$ and $\bar t$  caused by such a CP violation~\cite{asakawa}.
The large mass of $t$ implies that it decays before hadronization. As a result 
it acts as a heavy lepton and the spin information gets translated to 
distribution of the decay leptons. Thus we can use these angular distributions
as a probe  for possible CP violation.  We use only the decay lepton angular 
distributions and construct asymmetries that reflect the $t/ \bar t$ 
polarization asymmetries caused by the CP violation in the Higgs couplings. 
We observe that these are independent of any CP-violating contribution in 
the $tbW$ vertex. The same is not true of the energy distribution of the 
decay lepton. Hence we restrict our analysis to the angular distributions and 
keep the $tbW$ vertex completely general, choosing the $t \bar t \gamma$ vertex
to be standard. The latter of course is relevant only for the 
continuum background.\\\\
We develop a strategy to study the CP property of the Higgs by looking at 
angular distributions of leptons and antileptons, for different polarizations 
of  the colliding photons. Towards this end  we obtain analytical expressions 
for the lepton angular distribution, with a fixed value of the photon
energy and general polarization. We then fold this expression with the photon
luminosity function and the polarization profile  for the ideal back-scattered
laser spectrum \cite{ginzburg}; we obtain numerical results for the different 
mixed polarisation and charge asymmetries which we construct. 
Our choice of the ideal case for the back-scattered laser spectrum is for 
demonstration purposes. Further results using the recently available 
spectra, including the detector simulation for TESLA \cite{zarnecki,telnov12},
will be presented elsewhere \cite{usfuture}.  We then use the above-mentioned 
asymmetries to assess the sensitivity
of this process to the size of various form factors involved  in the
parametrization of ${\cal V}_{\gamma \gamma \phi}, {\cal V}_{t \bar t \phi}$.
At times we have used specific predictions for the form factors in the 
MSSM~\cite{asakawa} as a guide and for purposes of illustration,
in our analysis. We  show that this process is capable of probing
the MSSM loop effects using these asymmetries. \\\\
The plan of the paper is as follows. In section \ref{two} 
we give the general form for CP-violating vertices involving the
Higgs, the $t$ quark  and the photons as well as the $tbW$ decay vertex for 
the $t$ quark. The $t \bar t$ production and $t$-decay helicity amplitudes 
obtained with  these vertices are then presented. In section \ref{three} we 
obtain an analytic expression for the angular distribution of the decay 
lepton in  $t/\bar t$ decay.  We discuss the insensitivity of the decay-lepton 
angular distribution to the anomalous coupling in the $tbW$ vertex. Section
\ref{four} deals with the ideal photon collider \cite{ginzburg}. Numerical 
results are presented in section \ref{five}, discussed in section \ref{six}
and we  conclude in section \ref{seven}.\\\\

%
%
\section{Interaction Vertices and  \hspace{1cm}Helicity Amplitudes}
\label{two}
The interaction vertex of $t$ with a scalar $\phi$, which may or may not be a
CP eigenstate, may be written in a model-independent way as
\begin{equation}
{\cal V}_{t\bar t \phi} = -ie\frac{m_t}{M_W} \left(S_t+i\gamma^5P_t\right).
\label{v1}
\end{equation}
The general expression for the loop-induced $\gamma\gamma\phi$ vertex can
be parametrized as
\begin{eqnarray}
{\cal V}_{\gamma\gamma\phi}=\frac{-i\sqrt{s}\alpha}{4\pi}\left[S_{\gamma}(s)
\left(\epsilon_1.\epsilon_2-\frac{2}{s}(\epsilon_1.k_2)(\epsilon_2.k_1) \right)
\right. \nonumber\\
-\left. P_{\gamma}(s)\frac{2}{s}\epsilon_{\mu\nu\alpha\beta}\epsilon_1^{\mu}
\epsilon_2^{\nu} k_1^{\alpha} k_2^{\beta} \right],
\label{v2}
\end{eqnarray}
where $k_1$ and $k_2$ are the four-momenta of colliding photons and 
$\epsilon_{1,2}$ are corresponding polarization vectors. We take 
$S_\gamma, P_\gamma$ to be complex whereas $S_t,P_t$ are taken to be real.
This choice means that we assume only the CP mixing coming from the 
loop-induced effects in the Higgs potential. We allow these form factors to be 
slowly varying functions of the $\gamma \gamma$ c.m. energy since in any model 
the loop-induced couplings will have
such a dependence. Simultaneous non-zero values for $P$ and $S$ form factors
signal CP violation. We will construct various asymmetries which can
give information on these form factors.\\\\
We allow the $tbW$ vertex to be completely general and write it as
\begin{eqnarray}
\Gamma^{\mu}_{tbW}=-\frac{g}{\sqrt{2}}V_{tb}\left[\gamma^{\mu}
\left(f_{1L}P_L + f_{1R} P_R\right)\right.\nonumber\\
-\left.\frac{i}{M_W}\sigma^{\mu\nu}(p_t-p_b)_{\nu}
\left(f_{2L}P_L + f_{2R} P_R\right)\right],\\
{\bar\Gamma}^{\mu}_{\bar t\bar b W}=-\frac{g}{\sqrt{2}}V_{tb}^*\left[
\gamma^{\mu}\left({\bar f}_{1L}P_L + {\bar f}_{1R} P_R\right)\right.\nonumber\\
-\left.\frac{i}{M_W}\sigma^{\mu\nu}
(p_{\bar t}-p_{\bar b})_{\nu}
\left({\bar f}_{2L}P_L + {\bar f}_{2R} P_R\right)\right],
\end{eqnarray}
where $V_{tb}$ is the CKM matrix element and $g$ is the $SU(2)$ coupling.
We work in the approximation of vanishing $b$ mass. Hence $f_{1R}, {\bar f}_{1R}
,f_{2L}$ and ${\bar f}_{2R}$ do not contribute. We choose SM values for
$f_{1L}$, and ${\bar f}_{1L}$, viz.,  $f_{1L}
={\bar f}_{1L} =1 $. The only non-standard part of the $tbW$ vertex which gives
non-zero contribution then corresponds to the terms with 
$f_{2R}$ and ${\bar f}_{2L}$. We will sometimes also use the 
notation $f^+$ and $f^-$ for $f_{2R}$ and $\bar f_{2L}$ respectively.
One expects these unknown $f$'s to be small and we  retain only linear terms in
them while calculating the amplitudes. Below we give the helicity amplitudes for
the production of $t \bar t$ followed by the decays of the $t/\bar t$ in terms of these
general couplings.\\
\subsection{Production Helicity Amplitude}
The production process $\gamma \gamma \rightarrow t \bar t$ receives the $t/u$
channel SM contribution from the first two diagrams of Fig. 1, which is 
CP-conserving whereas the $s$ channel $\phi$ exchange contribution may be 
potentially CP violating. The helicity amplitudes for the $s$ and the 
$t/u$ channel diagrams are 
given by Eqs. (\ref{prodhig}) and (\ref{prodsm}) respectively :
\begin{eqnarray}
%
&& M_{\phi}(\lambda_1,\lambda_2;\lambda_t,\lambda_{\bar t})= \frac{-ie\alpha
 m_t}{4\pi M_W} \ \frac{s}{s-m_{\phi}^2+im_{\phi}\Gamma_{\phi}}\nonumber\\
&& \left[ S_{\gamma}(s) + i\lambda_1 P_{\gamma}(s)\right]
 \left[\lambda_t \beta S_t - iP_t\right] \ \delta_{\lambda_1,\lambda_2}\delta_
 {\lambda_t,\lambda_{\bar t}},
 \label{prodhig}\\
&& M_{SM}(\lambda_1,\lambda_2;\lambda_t,\lambda_{\bar t})= \frac{-
 i4\pi\alpha Q^2}{1-\beta^2 \cos^2\theta_t} \nonumber\\
&& \left[\frac{4m_t}{\sqrt{s}} \ (\lambda_1+\lambda_t\beta)
 \ \delta_{\lambda_1,\lambda_2}\delta_{\lambda_t,\lambda_{\bar t}}
 \right.\nonumber\\
&& -\frac{4m_t}{\sqrt{s}} \ \lambda_t\beta\sin^2\theta_t \ \delta_{\lambda_1,-
 \lambda_2}\delta_{\lambda_t,\lambda_{\bar t}}\nonumber\\
&& -\left. 2\beta \ (\cos\theta_t + \lambda_1\lambda_t) \sin\theta_t \ \delta_
 {\lambda_1,-\lambda_2}\delta_{\lambda_t,-\lambda_{\bar t}}\right].
 \label{prodsm}
\end{eqnarray}
Here $\beta, \ Q$ and $\theta_t$ are velocity, electric charge and scattering
angle of the $t$ quark respectively; $\Gamma_{\phi}$ and $m_{\phi}$ denote the 
total decay width and mass of the scalar $\phi$; $\lambda_{1,2}$ 
stand for helicities of two photons while the other $\lambda$'s stand for 
helicities of particles indicated by the subscript. For photons, helicities are
written in units of $\hbar$ while for spin-1/2 fermions they are in units of 
$\hbar/2$.\\\\
From the expressions in Eqs. (\ref{prodhig}) and (\ref{prodsm}) it is clear 
that  the $\phi$ exchange diagram contributes only when both colliding 
photons have the same helicities, whereas the SM contribution is small for 
this combination as we move away from the $t \bar t$ threshold. 
Thus a choice of equal helicities for both colliding photons can maximize 
polarization asymmetries for the produced $t \bar t$ pair, better reflecting 
the CP-violating nature of the $s$-channel contribution. It should be 
mentioned here that these statements are true only in the leading order (LO).
Radiative corrections to $\gamma \gamma \rightarrow q \bar q$ 
can be large \cite{jikia,melles}. That is also the reason we 
have restricted our analysis to asymmetries, which involve ratios. As a result
the analysis is quite robust even if we use only the LO result for the
SM contribution.  Note also that the SM contribution for equal photon
helicities is peaked 
in the forward and backward directions, whereas the scalar-exchange contribution
is independent of the production angle $\theta_t$.  This also
suggests that one can optimize the asymmetries by angular cuts to reduce the
SM contributions to the integrated cross-section, of course taking care that
the total event rate is not reduced too much. We will make use of this feature
in our studies.
\subsection{Decay Helicity Amplitudes}
We assume that the $t$ quark decays only via the $tbW$ vertex followed by the 
decay of $W$ into lepton and corresponding neutrino. The helicity amplitudes
$ M_{\Gamma}(\lambda_t,\lambda_b,\lambda_{l^+},\lambda_{\nu})$ and
$ \overline M_{\Gamma}(\lambda_{\bar t},\lambda_{\bar b},\lambda_{l^-},
\lambda_{\bar \nu})$, for the decay of $t$ and $\bar t$ are given below:
\begin{eqnarray}
M_{\Gamma}(+,-,+,-) = -i2\sqrt{2} g^2 \sqrt{m_t p^0_b p^0_{\nu} p^0_{l^+}} \
\Delta_W(p_W^2)\nonumber\\
 \times \left\{\left(1+\frac{f_{2R}}{\sqrt{r}}\right)\cos\frac{\theta_{l^+}}{2}
\left[ \cos\frac{\theta_{\nu}}{2}\sin\frac{\theta_{b}}{2}e^{i\phi_b} -
\right.\right.\nonumber\\
\left.\sin\frac{\theta_{\nu}}{2}\cos\frac{\theta_{b}}{2} e^{i\phi_{\nu}}\right]
 - \frac{f_{2R}}{\sqrt{r}}\sin\frac{\theta_{b}}{2}e^{i\phi_{b}}\nonumber\\
\left.\left[
\sin\frac{\theta_{\nu}}{2}\sin\frac{\theta_{l^+}}{2}e^{i(\phi_{\nu}-
\phi_{l^+})}+\cos\frac{\theta_{\nu}}{2}\cos\frac{\theta_{l^+}}{2}\right]
\right\},
\end{eqnarray}
\begin{eqnarray}
M_{\Gamma}(-,-,+,-) = -i2\sqrt{2} g^2 \sqrt{m_tp^0_b p^0_{\nu} p^0_{l^+}} \
\Delta_W(p_W^2) \nonumber\\
 \times \left\{\left(1+\frac{f_{2R}}{\sqrt{r}}\right)\sin\frac{\theta_{l^+}}{2}
e^{-i\phi_{l^+}} \left[ 
\cos\frac{\theta_{\nu}}{2}\sin\frac{\theta_{b}}{2}e^{i\phi_b}
\right.\right.\nonumber\\
-\left.\sin\frac{\theta_{\nu}}{2}\cos\frac{\theta_{b}}{2}e^{i\phi_n}\right]
 +\frac{f_{2R}}{\sqrt{r}}\cos\frac{\theta_{b}}{2}\nonumber\\
\left.\left[
\sin\frac{\theta_{\nu}}{2}\sin\frac{\theta_{l^+}}{2}e^{ i(\phi_{\nu}-
\phi_{l^+})}+\cos\frac{\theta_{\nu}}{2}\cos\frac{\theta_{l^+}}{2}\right]
\right\},
\end{eqnarray}
\begin{eqnarray}
\overline M_{\Gamma}(+,+,-,+) = -i2\sqrt{2} g^2 \sqrt{m_t p^0_{\bar b} p^0_{\bar\nu}
 p^0_{l^-}} \ \Delta_W(p_W^2)\nonumber\\
 \times \left\{\left(1+\frac{\bar 
f_{2L}}{\sqrt{r}}\right)\cos\frac{\theta_{l^-}}{2}
\left[\cos\frac{\theta_{\bar\nu}}{2}\sin\frac{\theta_{\bar 
b}}{2}e^{-i\phi_{\bar b}}
\right.\right.\nonumber\\
-\left.\sin\frac{\theta_{\bar\nu}}{2}\cos\frac{\theta_{\bar b}}{2}
e^{-i\phi_{\bar\nu}}\right]
 - \frac{\bar f_{2L}}{\sqrt{r}}\sin\frac{\theta_{\bar b}}{2}e^{-i\phi_{\bar
b}}\nonumber\\
\left.\left[\sin\frac{\theta_{\bar\nu}}{2}\sin\frac{\theta_{l^-}}{2}e^{-i(\phi_{\bar\nu}-
\phi_{l^-})}+\cos\frac{\theta_{\bar\nu}}{2}\cos\frac{\theta_{l^-}}{2}\right]
\right\},
\end{eqnarray}
\begin{eqnarray}
\overline M_{\Gamma}(-,+,-,+) = -i2\sqrt{2} g^2 \sqrt{m_tp^0_{\bar b} p^0_{\bar\nu}
 p^0_{l^-}} \ \Delta_W(p_W^2) \nonumber\\
 \times \left\{\left(1+\frac{\bar 
f_{2L}}{\sqrt{r}}\right)\sin\frac{\theta_{l^-}}{2}
e^{i\phi_{l^-}}
\left[\cos\frac{\theta_{\bar\nu}}{2}\sin\frac{\theta_{\bar 
b}}{2}e^{-i\phi_{\bar b}}
\right.\right.\nonumber\\
-\left.\sin\frac{\theta_{\bar\nu}}{2}\cos\frac{\theta_{\bar b}}{2}e^{-i\phi_
{\bar\nu}}\right]
 + \frac{\bar f_{2L}}{\sqrt{r}}\cos\frac{\theta_{\bar b}}{2}\nonumber\\
\left.\left[
\sin\frac{\theta_{\bar\nu}}{2}\sin\frac{\theta_{l^-}}{2}e^{-i(\phi_{\bar\nu}-
\phi_{l^-})}+\cos\frac{\theta_{\bar\nu}}{2}\cos\frac{\theta_{l^-}}{2}\right]
\right\},
\end{eqnarray}
where,
$$\Delta_W(p_W^2) \ = \ \frac{1}{p_W^2 - M_W^2 +i M_W \Gamma_W},\hspace{0.2cm}
r=\frac{M_W^2}{m_t^2}.$$
For simplicity, the above expressions for the decay amplitudes have been 
calculated in the rest frame of $t$ ($\bar t$) with the $z$-axis pointing in 
the direction of its momentum in the $\gamma\gamma$ c.m. frame. We treat the 
decay lepton $l$ and the $b$ quark as massless and list only the non-zero
amplitudes.
%
%
\section{Angular Distribution of Leptons}
\label{three}
Using the narrow-width approximation for the $t$ quark and the $W$ boson, the
differential cross section for $\gamma \gamma \rightarrow t \bar t
\rightarrow l^+ b \nu_l \bar t$ can be written in terms of the density matrices 
as 
\begin{eqnarray}
\frac{d\sigma}{d\cos\theta_t \ d\cos\theta_{l^+} \ dE_{l^+} \ d\phi_{l^+}}=
\frac{3e^4 g^4 \beta E_{l^+}}{64(4\pi)^4 s \Gamma_tm_t \Gamma_WM_W}
\nonumber\\
\ \sum_{\lambda,\lambda'} \underbrace{\rho^{'+}(\lambda,\lambda')}_{\rm c.m. \ 
frame} \underbrace{\left[\frac{\Gamma'(\lambda,\lambda')}{m_t E_{l^+}^0}
\right]}_{\rm rest \ frame}.\nonumber\\
\label{sigden}
\end{eqnarray}
Here $E_{l^+}^0$ is the energy of $l^+$ in the rest frame of the $t$ quark; the
production and decay density matrices are given by
\begin{eqnarray*}
\rho^+(\lambda,\lambda')&=&e^4\rho^{'+}(\lambda,\lambda') = \sum
\rho_1(\lambda_1,\lambda_1')\rho_2(\lambda_2,\lambda_2') \\
&&\times \ {\cal M}(\lambda_1,\lambda_2,\lambda,\lambda_{\bar t})
{\cal M}^*(\lambda_1',\lambda_2', \lambda',\lambda_{\bar t})\\[0.2cm]
{\rm and}&&\\
\Gamma(\lambda,\lambda')&=&g^4 |\Delta(p_W^2)|^2  \ \Gamma'(\lambda,
\lambda') = \frac{1}{2\pi}\int d\alpha \\ 
&&\times \sum
M_{\Gamma}(\lambda,\lambda_b,\lambda_{l^+},\lambda_{\nu}) \
M_{\Gamma}^*(\lambda',\lambda_b,\lambda_{l^+},\lambda_{\nu}).
\end{eqnarray*}
Here $\alpha$ is the azimuthal angle of the $b$ quark in the rest frame of
the $t$ quark with the $z$-axis pointing in the direction of the momentum of 
the lepton.
All repeated indices of matrix elements and density matrices are summed
over; $\rho_{1(2)}$ are the photon density matrices;  
written, in terms of the  Stokes parameters $\eta_i, \xi_i$:
\begin{eqnarray}
\rho_1(\lambda_1,\lambda_1')=\frac{1}{2}\left[ \begin{tabular}{cc}
$1+\eta_2$      &       $-\eta_3+i\eta_1$\\
$-\eta_3-i\eta_1$  &    $1-\eta_2$ \end{tabular}\right],\\
\rho_2(\lambda_2,\lambda_2')=\frac{1}{2}\left[ \begin{tabular}{cc}
$1+\xi_2$      &       $-\xi_3+i\xi_1$\\
$-\xi_3-i\xi_1$  &    $1-\xi_2$ \end{tabular}\right].
\end{eqnarray}
Here, $\eta_2$ is the degree of circular polarization while $\eta_1$ and $\eta_3$
are degrees of linear polarizations in two transverse directions of one photon;
$\xi_i$ are similarly the degrees of polarization for the second photon.
The explicit expressions for the production density matrix 
$\rho(\lambda_t,\lambda_t')$ depend upon the
polarization of initial photons. The decay density matrix
elements are independent of any initial condition and in the {\em rest frame} of
$t$ ($\bar t$) they are given by:
\begin{eqnarray}
\Gamma(\pm,\pm) = g^4m_tE_{l^+}^0|\Delta_W(p_W^2)|^2 \ (m_t^2-2p_t.p_{l^+
})\nonumber\\
(1\pm\cos\theta_{l^+})\left(1+\frac{\Re(f_{2R})}{\sqrt{r}}
\frac{M_W^2}{p_t.p_{l^+}}\right),\label{Gtpp}\\
\Gamma(\pm,\mp) = g^4m_tE_{l^+}^0|\Delta_W(p_W^2)|^2 \ (m_t^2-2p_t.p_{l^+
})\nonumber\\
\sin\theta_{l^+}e^{\pm i\phi_{l^+}}\left(1+\frac{\Re(f_{2R})}{\sqrt{r}}
\frac{M_W^2}{p_t.p_{l^+}}\right),\label{Gtpm}\\
\overline \Gamma(\pm,\pm) = g^4m_tE_{l^-}^0|\Delta_W(p_W^2)|^2 \ (m_t^2-2p_{\bar
t}.p_{l^-})\nonumber\\
(1\pm\cos\theta_{l^-})\left(1+\frac{\Re(\bar f_{2L})}{\sqrt{r}}
\frac{M_W^2}{p_{\bar t}.p_{l^-}}\right),\label{GTpp}\\
\overline \Gamma(\pm,\mp) = g^4m_tE_{l^-}^0|\Delta_W(p_W^2)|^2 \ (m_t^2-2p_{\bar
t}.p_{l^-})\nonumber\\
\sin\theta_{l^-}e^{\mp i\phi_{l^-}}\left(1+\frac{\Re(\bar f_{2L})}
{\sqrt{r}}\frac{M_W^2}{p_{\bar t}.p_{l^-}}\right).\label{GTpm}
\end{eqnarray}
We have kept only the linear terms in the form factors $f_{2R}$ and
$\bar f_{2L}$, as we assume them to be small. These expressions written in terms
of the lab variables can also be written in terms of the variables in the
$\gamma\gamma$ c.m. frame. The relations between the angles in the rest frame 
and the $\gamma\gamma$ c.m. frame can be easily derived and are given by
\begin{eqnarray}
&&1\pm\cos\theta_{l^+} = \frac{(1\mp\beta)(1\pm\cos\theta_{tl^+}^{c.m.})}{1-
\beta\cos\theta_{tl^+}^{c.m.}},\\
&&1\pm\cos\theta_{l^-} = \frac{(1\pm\beta)(1\mp\cos\theta_{\bar tl^-}^{c.m.})}{1-
\beta\cos\theta_{\bar tl^-}^{c.m.}},\\
&&\sin\theta_{l^+}e^{i\phi_{l^+}} = \frac{\sqrt{1-\beta^2}}{1-\beta\cos\theta_
{tl^+}^{c.m.}}\nonumber\\
&&(\sin\theta_{l^+}^{c.m.}\cos\theta_t^{c.m.}\cos\phi_{l^+}^{c.m.}-
\cos\theta_{l^+}^{c.m.}\sin\theta_t^{c.m.}\nonumber\\
&&+i\sin\theta_{l^+}^{c.m.}\sin\phi_{l^+}^{c.m.}),\\
&&\sin\theta_{l^-}e^{i\phi_{l^-}} = \frac{\sqrt{1-\beta^2}}{1-\beta\cos\theta_
{\bar tl^-}^{c.m.}}\nonumber\\
&&(-\sin\theta_{l^-}^{c.m.}\cos\theta_{\bar t}^{c.m.}\cos\phi_{l^-}^{c.m.}+
\cos\theta_{l^-}^{c.m.}\sin\theta_{\bar t}^{c.m.}\nonumber\\
&&+i\sin\theta_{l^-}^{c.m.}\sin\phi_{l^-}^{c.m.}).
\end{eqnarray}
Using the above relations and dropping the superscripts {\em c.m.} from the 
angles, we can rewrite Eq. (\ref{sigden}) as
\begin{eqnarray}
&&\frac{d\sigma}{d\cos\theta_td\cos\theta_{l^{\pm}}dE_{l^{\pm}}d\phi_
{l^{\pm}}}=
\frac{3\alpha^4\beta}{16x_w^2\sqrt{s}}\frac{E_{l^{\pm}}}{\Gamma_t\Gamma_WM_W}
\nonumber\\
&&\frac{1}{8\gamma_t}
\left(\frac{1}{1-\beta\cos\theta_{tl}} -\frac{4E_{l^{\pm}}}{\sqrt{s}(1-\beta^2)}\right)\nonumber\\
&&\times\left(1+\frac{\Re(f^{\pm})}{\sqrt{r}}\frac{2M_
W^2}{E_{l^\pm}\sqrt{s}(1-\beta\cos\theta_{tl})}\right)\nonumber\\\nonumber\\
&&\times\left[A^{\pm}(1-\beta\cos\theta_{tl}) \pm B^{\pm}(\cos\theta_{tl}-\beta)\right.
\nonumber\\
&&\pm C^{\pm}\sqrt{1-\beta^2}\sin\theta_{l^{\pm}}(\cos\theta_t
\cos\phi_{l^{\pm}}-\sin\theta_t\cot\theta_{l^{\pm}})\nonumber\\
&&\left. \pm D^{\pm}\sqrt{1-\beta^2}\sin\theta_{l^{\pm}}
\sin\phi_{l^{\pm}}\right],
\label{sigABC}
\end{eqnarray}
where $\gamma_t=1/\sqrt{1-\beta^2}$, $x_w=\sin^2\theta_W$, $\theta_W$ 
being the Weinberg angle and
\begin{eqnarray}
A^{\pm}&=&\rho^{'\pm}(+,+) + \rho^{'\pm}(-,-),\\
B^{\pm}&=&\rho^{'\pm}(+,+) - \rho^{'\pm}(-,-),\\
C^{\pm}&=&2\Re[\rho^{'\pm}(+,-)],\\
D^{\pm}&=&-2\Im[\rho^{'\pm}(+,-)].
\end{eqnarray}
In Eq. (\ref{sigABC}) and what follows, the lepton variables are defined
in the {\em $\gamma\gamma$ c.m. frame}. Further, $\theta_{tl}$
stands for $\theta_{t l^+}$ for the $l^+$ distribution and hence the upper
sign in the equation, whereas it stands for $\theta_{\bar t l^-}$ for the 
lower sign and hence the $l^-$ distribution.
To get the angular distribution of leptons we still have to integrate 
Eq. (\ref{sigABC}) over $E_l, \ \cos\theta_t$ and $\phi_l$. 
The limits on $E_l$ integration are,
\begin{eqnarray*}
\frac{M_W^2}{\sqrt{s}} \ \frac{1}{1-\beta\cos\theta_{tl}} \ \leq \ E_l \ \leq
\frac{m_t^2}{\sqrt{s}} \ \frac{1}{1-\beta\cos\theta_{tl}}.
\end{eqnarray*}
After the $E_l$ integration, we get
\begin{eqnarray}
&&\frac{d\sigma}{d\cos\theta_t \ d\cos\theta_{l^\pm} \ d\phi_{l^\pm}}
 = \frac{3\alpha^4\beta}{16x_w^2\sqrt{s}} \ \frac{1}{\Gamma_t\Gamma_WM_W}
\nonumber\\
&&\frac{1}{8\gamma_t} \ \frac{m_t^4}{6s} \ \frac{(1+2r-6\Re(f^\pm)\sqrt{r})
(1-r)^2}{(1-\beta\cos\theta_{tl})^3}\nonumber\\ \nonumber\\
&&\times \left[A^{\pm}(1-\beta\cos\theta_{tl}) \pm B^{\pm}(\cos\theta_{tl}
-\beta)\right.
\nonumber\\
&&\pm C^{\pm}\sqrt{1-\beta^2}\sin\theta_{l^{\pm}}(\cos\theta_t
\cos\phi_{l^{\pm}}-\sin\theta_t\cot\theta_{l^{\pm}})\nonumber\\
&&\left. \pm D^{\pm}\sqrt{1-\beta^2}\sin\theta_{l^{\pm}}
\sin\phi_{l^{\pm}}\right].
\label{sig3}
\end{eqnarray}
Here we have used the notation $f^+$  and $f^-$ for $f_{2R}$ and 
$\bar f_{2L}$, respectively. From the above equation it is clear that 
the angular distribution of leptons after
energy integration is modified due to the anomalous $tbW$ coupling only up to an
overall factor $1+2r-6\Re(f^\pm)\sqrt{r}$, which is independent of any 
kinematical variable. In fact, the same factor appears in the total width of 
the $t$ quark calculated up to linear order in $f^\pm$:
\begin{equation}
\Gamma_{t(\bar t)}  =  \frac{\alpha^2}{192x_w^2}  \frac{m_t^3}{\Gamma_WM_W}
 (1-r)^2\left[1+2r-6\Re(f^\pm)\sqrt{r}\right]
\end{equation}
and thus exactly cancels the one in Eq. (\ref{sig3}). Thus we see that
the angular distribution of the decay lepton is unaltered,  in the 
linear approximation of anomalous $tbW$ couplings.  In fact this is quite 
a general result, which is attained under certain assumptions and 
approximations we have made. We elaborate on this point below.\\\\
An inspection of Eqs. (\ref{Gtpp})--(\ref{GTpm}) shows that the presence of any
anomalous part in the $tbW$ coupling changes the decay density matrix only by
an overall  energy-dependent factor {\it independent} of angle. The quantity 
$p_t.p_l$ does have an apparent dependence on the angular variables of the 
lepton. However, in fact it depends only on the lepton energy. To see this 
clearly, let us go to the rest frame of the $t$ quark.
Now the three lepton variables are $\{E_l^{\rm rest},\theta_l^{\rm rest},
\phi_l^{\rm rest}\}$ and the anomalous term depends only on $E_l^{\rm rest}$. 
This means that the angular distribution of leptons in the rest frame of the $t$
quark is unaltered by the presence of an anomalous term in $tbW$ coupling, apart 
from an overall scaling. The angular distribution in any other frame can be 
obtained from that in the rest frame by a Lorentz boost. Thus the angular 
distribution of leptons in an arbitrary frame will be the same as that 
in the absence of the anomalous term, up to some overall factor that depends 
upon energy and the boost parameters and no angular variables. Of course, 
it is not very obvious by looking at Eq. (21) that this will indeed happen. 
But, with a change of variable,
$$E_l \rightarrow E_l^{\rm rest} = \gamma_tE_l(1-\beta\cos\theta_{tl}), $$
the additional overall factor becomes
$$1+\frac{\Re(f^\pm)}{\sqrt{r}}\frac{M_W^2}{m_tE_l^{\rm rest}},$$
which is clearly independent of angular variables. After integration 
over $E_l^{\rm rest}$, in the limit 
$M_W^2/2m_t \leq E_l^{\rm rest} \leq m_t/2 $ we get back 
Eq. (\ref{sig3}). The important point is that in proving the result we did not 
make any reference to the production density matrix and hence the result is 
very general and applicable to any $2\rightarrow 2$ process for $t \bar t$ pair
production provided the following conditions are fulfilled:
\begin{itemize}
\item we use the narrow-width approximation for $t$ and $W$,
\item $b, \ l, \ \nu$ are taken to be massless,
\item the only decay mode of $t$ is $t\rightarrow bW \rightarrow bl\nu $, and
\item the anomalous $tbW$ coupling is small enough that one can work to linear
approximation in it.
\end{itemize}
For the case of $e^+e^- \rightarrow t \bar t$ followed by subsequent
$t/\bar t$ decay,  this was observed  earlier \cite{sdr,hioki}.  It was proved
recently by two groups independently; for a two-photon initial state by 
Ohkuma~\cite{ohk}, for an arbitrary two-body initial state in \cite{grza} 
and further keeping $m_b$ non-zero in \cite{grza1}.  These
derivations use the method developed by Tsai and collaborators \cite{tsai} 
for incorporating the production and decay of a massive spin-half particle. 
Our derivation makes use of helicity amplitudes and provides an independent 
verification of these results. The result is very crucial for
our present work as we now have an observable where the only source of the 
CP-violating asymmetry will be the production process. \\\\
Thus we can analyse the Higgs CP property easily, as long as the anomalous part 
of the $tbW$ couplings, $f^\pm$ is small and the quadratic term can be 
neglected. If $f^\pm$ is not small then we have to keep the quadratic terms in 
Eqs. (\ref{Gtpp})-(\ref{Gtpm}) and the decay density matrices to this order are
then given by \footnote{Expressions for $\Gamma$ are written in the rest 
frame of the $t$ quark.}:
\begin{eqnarray}
\Gamma(\pm,\pm)=g^4m_tE_{l^+}^0|\Delta_W(p_W^2)|^2 \ (m_t^2-2p_t.p_{l^+
})\nonumber\\
\left[(1\pm\cos\theta_{l^+})\left(1+\frac{\Re(f^+)}{\sqrt{r}}
\frac{M_W^2}{p_t.p_{l^+}}\right)\right.\nonumber\\
-|f^+|^2(1\mp\cos\theta_{l^+})\left(1-
\frac{m_t^2+M_W^2}{2p_t.p_{l^+}}\right)\nonumber\\
\left.+|f^+|^2\frac{M_W^4}{2 r( p_t.p_{l^+})^2}
\cos\theta_{l^+}\right] ,\label{enGtpp}\\
\Gamma(\pm,\mp) = g^4m_tE_{l^+}^0|\Delta_W(p_W^2)|^2 \ (m_t^2-2p_t.p_{l^+
}) \nonumber\\
\sin\theta_{l^+}e^{\pm i\phi_{l^+}}
\left[1+\frac{\Re(f^+)}{\sqrt{r}}\frac{M_W^2}{p_t.p_{l^+}} \right.\nonumber\\
+ \left. |f^+|^2
\left( 1-\frac{m_t^2+M_W^2}{2p_t.p_{l^+}} + \frac{M_W^4}{2 r ( p_t.p_{l^+})^2}
\right)\right]. \label{enGtpm}
\end{eqnarray}
$\overline \Gamma(\lambda,\lambda')$ will be given by similar expressions. 
Thus if $f^\pm$ are not small they can modify the angular dependence of the 
decay 
density matrix in the rest frame of the $t$ quark and hence in any other frame. 
In that case it will not be trivial to use angular distributions to study the CP 
property of the production process. In this work we will assume $f^\pm$ to be 
small and will neglect the quadratic terms in Eqs. (\ref{enGtpp}) and 
(\ref{enGtpm}).\\\\
Making the above-mentioned four assumptions, which are very reasonable indeed, 
we now go on to calculate the final angular distribution by integrating 
Eq. (\ref{sig3}) over $\cos\theta_t$ and $\phi_l$. We obtain for the angular
distribution:
\begin{eqnarray}
&&\frac{d\sigma}{d\cos\theta_{l^\pm}} = \frac{3\pi\alpha^2\beta}{8s\gamma_t^2} \
\left[2A_{00}^{\pm}(I_{300}-\beta yI_{301})\right.\nonumber\\
&&+2A_{01}^{\pm}(I_{310}-\beta yI_{311})
+2A_{02}^{\pm}(I_{320} -\beta y I_{321})\nonumber\\
&&+2A_{22}^{\pm}(I_{322}-\beta yI_{323})+
2A_{42}^{\pm}(I_{324} -\beta yI_{325})\nonumber\\
&&\pm\frac{B_{01}^{\pm}}{\beta}\{-2(I_{310}-\beta yI_{311})
+(1-\beta^2)\sum_{i=0}^{2}X_iI_{51i}\}\nonumber\\
&&\pm\frac{B_{02}^{\pm}}{\beta}\{-2(I_{320}-\beta yI_{321})
+(1-\beta^2)\sum_{i=0}^{2}X_iI_{52i}\}\nonumber\\
&&\pm\frac{B_{12}^{\pm}}{\beta}\{-2(I_{321}-\beta yI_{322})
+(1-\beta^2)\sum_{i=1}^{3}X_iI_{52i}\}\nonumber\\
&&\pm\frac{B_{32}^{\pm}}{\beta}\{-2(I_{323}-\beta yI_{324})
+(1-\beta^2)\sum_{i=3}^{5}X_iI_{52i}\}\nonumber\\
&&\left. \pm C_0 \ \sum_{j=0}^6 \ Y_j \ I_{52j}\right],
\label{sigang}
\end{eqnarray}
where
$$I_{ijk}=\intop^1_{-1}\frac{d\cos\theta_t\cos^k\theta_t}{[a+(y-
\beta\cos\theta_t)^2]^{i/2}(1-\beta^2\cos^2\theta_t)^j},$$
\begin{tabular}{ccl}
$a$ & = & $(1-\beta^2)\sin^2\theta_{l^\pm}$,\\
$y$ & = & $\cos\theta_{l^\pm}.$
\end{tabular}\\\\
We have obtained explicit analytical expressions for all $I_{ijk}$ 
appearing in Eq. (\ref{sigang}), which  are not listed here.
$A_{ij}$ and $B_{ij}$ are coefficients in expansions of the following type:
$$A^{\pm}=\sum_{i,j}\frac{A^{\pm}_{ij}\cos^i\theta_t}{(1-\beta^2\cos^2\theta_
t)^j}\hspace{1cm}{\rm etc.} $$
Expressions for $A_{ij}$'s and $B_{ij}$'s for circular polarization of photons
and expressions for $X_i$'s, $Y_j$'s and $C_0$ are given in the appendix.
Equation (\ref{sigang}) is the angular distribution of leptons for a given $\gamma
\gamma$ centre-of-mass energy. In a $\gamma\gamma$-collider 
constructed using the back-scattered laser beam one will not have monoenergetic
photons in the inital state; further, the degree of circular polarization 
of the photons will depend on its energy.  Thus the final observable 
cross section is to be obtained by folding Eq. (\ref{sigang}) with the 
luminosity function after accounting for the energy dependence of the 
circular polarization of the photons.
%
%
\section{Photon Collider}
\label{four}
\begin{figure}
\scalebox{1.00}{
\includegraphics{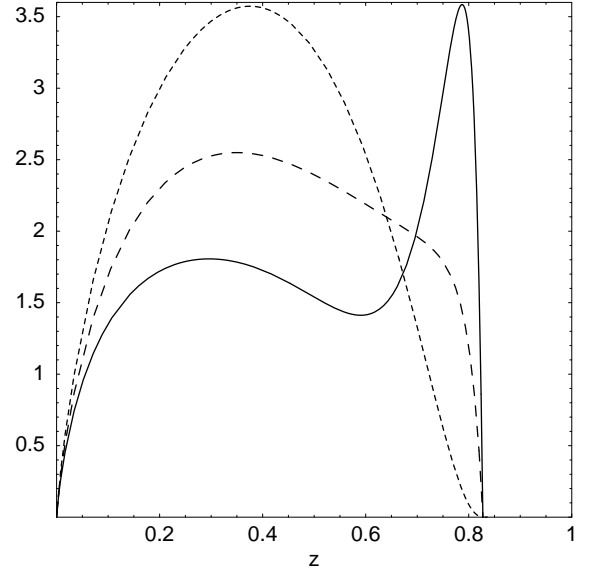}}
\caption{\label{spec}
Luminosity distribution  plotted against $z$ (which is related to the 
$\gamma\gamma$ invariant mass $W=2\sqrt{\omega_1\omega_2}$ by $z=W/(2E_b)$)
for $x=4.8$. Solid line corresponds to $\lambda_e\lambda_l=-1$, small dashed 
line is for $\lambda_e\lambda_l=1$ and large dashed line is for $\lambda_e 
\lambda_l=0$. Conversion distance is taken to be zero.}
\end{figure}

\begin{figure}
\scalebox{1.00}{
\includegraphics{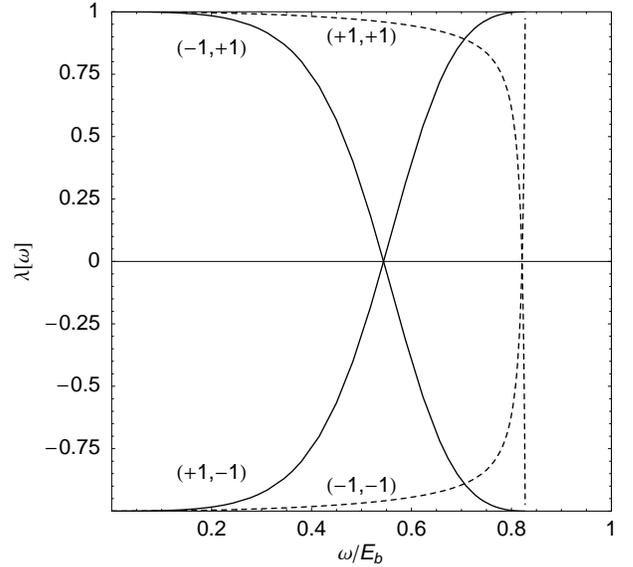}}
\caption{\label{pol}
Mean helicity of scattered high energy photon plotted against reduced
energy of photon $y=\omega/E_b$. 
Solid lines are for $\lambda_e\lambda_l=-1$ and
the dashed lines corresponds to $\lambda_e\lambda_l=1$. The lines are marked
with the values of ($\lambda_e,\lambda_l$).}
\end{figure}

In a $\gamma\gamma$ collider, high energy photons are produced by Compton
back-scattering of a laser from high energy $e^-$ or $e^+$ beam  via
\begin{eqnarray*}
e^-(\lambda_{e^-}) \ \gamma(\lambda_{l_1}) \rightarrow e^- \ 
\gamma(\lambda_1)\\
e^+(\lambda_{e^+}) \ \gamma(\lambda_{l_2}) \rightarrow e^+ \ \gamma(\lambda_2).
\end{eqnarray*}
In this paper we will be using the ideal photon spectrum due to Ginzburg {\it
et al.} for $x \leq 4.8$. The ideal luminosity (for zero conversion distance)
is given by 
\begin{equation}
\frac{1}{L_{e^-e^+}}\frac{dL_{\gamma\gamma}}{dy_1dy_2}=f(y_1)f(y_2),
\label{lumin}
\end{equation}
where
\begin{eqnarray}
f(y)&=&\frac{2\pi\alpha^2}{\sigma_c x m_e^2}\left[\frac{1}{1-y}+1-y-4r(1-r)
\right. \nonumber\\
&-&\left. \lambda_e \lambda_l r x (2r-1)(2-y)\right],\\
x&=&\frac{4E_b \omega_0}{m_e^2}=15.3\left(\frac{E_b}{\rm TeV}\right) \ \left(
\frac{\omega_0}{\rm eV}\right),\\
r&=&\frac{y}{x(1-y)}\leq 1,\\
y&=&\frac{\omega}{E_b},\hspace{2cm}\omega \leq \frac{xE_b}{1+x},\\
\sigma_c&=&\sigma_c^{np}+\lambda_e\lambda_l\sigma_1,\\
\sigma_c^{np}&=&\frac{2\pi\alpha^2}{xm_e^2}\left[\left(1-\frac{4}{x}-
\frac{8}{x^2}\right)\log(1+x)\right. \nonumber\\
&+&\left.\frac{1}{2}+\frac{8}{x}- \frac{1}{2(1+x)^2}
\right],\\
\sigma_1&=&\frac{2\pi\alpha^2}{xm_e^2}\left[\left(1+\frac{2}{x}\right) 
\log(1+x)
\right. \nonumber\\
&-&\left.\frac{5}{2}+\frac{1}{1+x}-\frac{1}{2(1+x)^2} \right],
\end{eqnarray}
$\lambda_e$ and $\lambda_l$ are the initial electron (positron) and laser
helicities respectively, and $\omega_0$ is the energy of the laser.
In Eq. (\ref{lumin}), if we change variables from $y_1$ and $y_2$ to
$z=\sqrt{y_1y_2}$ and $y_2$ and integrate over $y_2$, we will get an expression
for the photon spectrum as a function of the $\gamma\gamma$ invariant mass 
$W= 2zE_b$, where $E_b$ is the energy of the $e^{\pm}$ beam and is plotted in 
Fig. \ref{spec} for $x=4.8$. The spectrum is peaked in the hard photon region 
for $\lambda_e \lambda_l=-1$.\\\\
The mean helicity of high energy photons depends upon their energy in the
\noindent
lab frame. In an ideal $\gamma\gamma$
collider the energy dependence of the mean helicity is given by
\begin{eqnarray}
\eta_2(y)&=&\frac{2\pi\alpha^2}{\sigma_c x m_e^2 f(y)}
\left\{\lambda_l(1-2r)(1-y+1/(1-y)) \right.\nonumber\\
&&+\left.\lambda_e r x [1+(1-y)(1-2r)^2]\right\}
\end{eqnarray}
and is plotted in Fig. \ref{pol} for $x=4.8$.
\noindent
For $\lambda_e\lambda_l=-1$ the back-scattered photon has the same helicity as 
the electron (positron). Also, the spectrum is peaked at high energy, and yields
a high degree of polarisation of the photon beam. Hence, the dominant photon
polarization in this case is decided by the electron (positron) helicity.
Now, as suggested by Eq. (\ref{prodhig}), the helicities of two colliding 
photons should be equal in order to have Higgs contribution. Thus we choose 
$\lambda_e\lambda_l=-1$ to get a hard photon spectrum, 
and set $\lambda_{e^-} = \lambda_{e^+}$ to maximize
the Higgs contribution, and hence the sensitivity to possible CP-violating
interactions. For our numerical analysis, we have chosen 
$\lambda_e\lambda_l=-1$; the initial state can thus be completely described 
by the helicities of the initial  electron and positron. We denote the total 
cross section in the lab frame by $\sigma (\lambda_{e^-},\pm)$, where the 
second argument denotes the charge of the final state lepton.\\\\
The total cross-section with an angular cut in the lab frame can be obtained by
folding Eq. (\ref{sigang}) with the photon spectrum:
\begin{eqnarray}
\sigma(\lambda_{e^-},\pm)=\intop \!\!\frac{dL_{\gamma\gamma}} {dy_1
dy_2}dy_1 dy_2\intop_{f(\theta_0,y_1,y_2)}^{g(\theta_0,y_1,y_2)}\!\!\!\!
d\cos\theta_{l^{\pm}} \frac{d\sigma}{d\cos\theta_{l^{\pm}}},\nonumber\\
\label{fold}
\end{eqnarray}
where $f(\theta_0,y_1,y_2)$ and $g(\theta_0,y_1,y_2)$ are the (boosted) limits
on integration in the $\gamma\gamma$ c.m. frame.
We end this section  with a few remarks. We have presented the specific case
where the $\gamma \gamma$ collider is based on a parent $e^+e^-$ collider
and we also assume $100 \%$ polarization for the $e^-/e^+$.  The analysis 
is completely valid for the case of a parent $e^- e^-$ collider, 
for which achieving a high degree of polarization for initial leptons might 
be technologically simpler.
%
%
\section{Numerical Results}
\label{five}
To determine the CP properties of the Higgs, we need to know all the four
form factors appearing in Eqs. (\ref{v1}) and (\ref{v2}). Assuming the mass 
and decay width of Higgs to be known, we then have the following 
six unknowns
$$S_t, \ P_t, \ \Re(S_{\gamma}), \ \Im(S_{\gamma}), \ \Re(P_{\gamma}), \
\Im(P_{\gamma}). $$
They appear in eight combinations in the expression for the production density
matrix, which we denote by $x_i$ and $y_i$ $(i=1,..., 4)$,
and are listed below, together with their CP properties.
\begin{center}
\begin{tabular}{ccc}\hline
Combinations & Aliases & CP property\\ \hline
$S_t\Re(S_{\gamma})$ & $x_1$ &even\\
$S_t\Im(S_{\gamma})$ & $x_2$ &even\\
$S_t\Re(P_{\gamma})$ & $y_1$ &odd\\
$S_t\Im(P_{\gamma})$ & $y_2$ &odd\\
$P_t\Re(S_{\gamma})$ & $y_3$ &odd\\
$P_t\Im(S_{\gamma})$ & $y_4$ &odd\\
$P_t\Re(P_{\gamma})$ & $x_3$ &even\\
$P_t\Im(P_{\gamma})$ & $x_4$ &even\\ \hline
\end{tabular}
\end{center}
Only five of these combinations are independent because they satisfy the 
following relations
\begin{eqnarray*}
y_1 \ . \ y_3 \ = \ x_1 \ . \ x_3, \hspace{1cm} y_2 \ . \ y_4 \ = \ x_2 \ . \
x_4, \\
y_1 \ . \ x_4 \ = \ y_2 \ . \ x_3, \hspace{1cm} y_4 \ . \ x_1 \ = \ y_3 \ . \
x_2.
\end{eqnarray*}
Any three relations of the four listed above are independent relations while
the fourth one is derived. Expressions for asymmetries can by written in terms
of $x$'s and $y$'s and can be used to put limits on sizes on these combinations.
\\\\
In what follows we will define various asymmetries involving the
polarization of the initial $e$ and charge of the final decay lepton, 
some of which are  CP-violating and use them to put limits on the size 
of various combinations of the form factors. There is no forward--backward 
asymmetry because two photons with the same helicities are indistinguishable in 
their c.m. frame. That is, no favoured direction exists and the forward 
direction is indistinguishable from the backward. This is to be contrasted with
the situation studied in \cite{pou}, where forward--backward asymmetry could be
used to put limits on CP violation arising from the top electric dipole
moment or a CP-odd $\gamma\gamma Z$ coupling.  The effects 
of the $s$-channel Higgs-exchange diagram appear only in charge and
polarization asymmetries along with purely CP-violating asymmetries.
For our numerical studies we take the values of the form factors calculated
in the MSSM for certain values of its parameters.  The specific 
values which we use for demonstration purposes
are taken from Ref. \cite{asakawa}, These are for $\tan \beta =3$, with all 
sparticles heavy and maximal phase:
\begin{eqnarray}
\begin{tabular}{cclcccl}
$m_\phi$&=& $500$  GeV &,& $\Gamma_\phi$&=& $1.9$  GeV,\\ 
$S_t$   &=& $ 0.33    $ &,& $P_t$  &=& $0.15$,\\
$S_{\gamma}$&=& $-1.3-1.2i $&,&$ P_{\gamma}$&=&$-0.51+1.1i$.
\end{tabular}
\label{mssmparam}
\end{eqnarray}
For the SM, $S$'s and $P$'s are identically zero. By SM we mean contribution
only from $t$ and $u$ channels. The light CP-even Higgs contribution at the 
$t\bar t$ threshold and beyond is small and hence is neglected.
\subsection{Polarized Cross Sections and Asymmetries}
\begin{figure}
\includegraphics{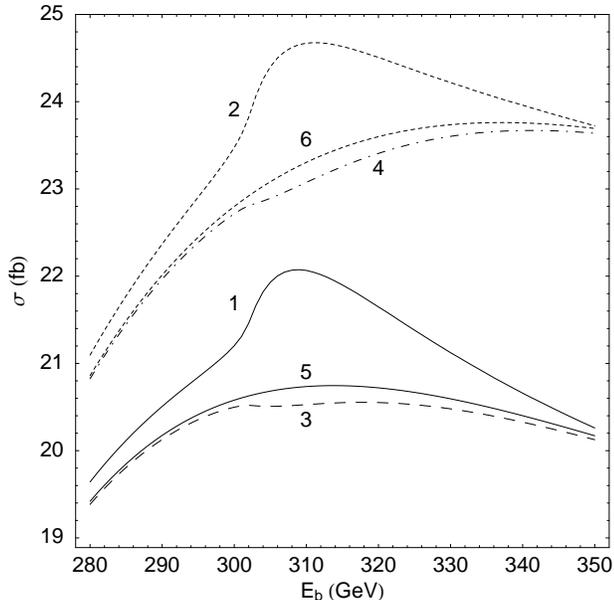}
\caption{\label{tSig}
All four integrated cross sections are plotted against the beam electron 
energy $E_b$, for the SM as well as for MSSM 
with our choice of parameters. The angular cut used in this 
figure is $60^{\circ}$ in the lab frame. Line 1 is for $\sigma(+,+)$, 
line 2  for $\sigma(+,-)$, line 3 for $\sigma(-,-)$ and 
line 4  for $\sigma(-,+)$ in the MSSM. Line 5 is for $\sigma(+,+)$ and
$\sigma(-,-)$ and line 6 for $\sigma(+,-)$ and $\sigma(-,+)$ in the SM.}
\end{figure}

There are two possibilities for the initial state polarization, $\lambda_{e^-}=
\lambda_{e^+} = +1 $ and $-1$. In the final state we can look for
either $l^+$ or $l^-$. This makes four possible polarized cross sections listed
as
$$\sigma(+,+), \ \sigma(+,-), \ \sigma(-,+), \ \sigma(-,-).$$
These are plotted in Fig.~\ref{tSig} as a function of the electron 
beam energy $E_b$ for
the angular cut of $60^{\circ}$ in the lab frame. For the SM, $\sigma(+,+)$ and
$\sigma(-,-)$ are exactly equal  as they are CP conjugates  of each other.
In the MSSM, because of CP violation, they can  differ. A similar statement
can be made about the pair $\sigma(-,+), \sigma(+,-)$. The flat behaviour 
with energy of curves $3$ and $4$ is due to the destructive interference
of the Higgs-mediated amplitude with the continuum. Recall here again that 
second index in the expressions of the cross sections is the sign of 
the charge of the lepton.  A comparison of curves $1,3$ and $5$, then shows 
clearly the change in the sign of interference effects as the sign of 
polarizations of the two photons is changed from $++$ to $--$.
The jump in $\sigma(+,+)$ and $\sigma(+,-)$ at around 310
GeV corresponds to matching of Higgs resonance peak with the peak of the hard
photon spectrum. This suggested to us the choice of $E_b =$ 310  GeV
for the analysis,
as the deviation from the SM is then very large for the chosen value of 
parameters.\\\\
We choose two polarized cross sections at a time,
out of the four available, and define six asymmetries as
\begin{eqnarray}
\mathcal{A}_1 & = & \frac{\sigma(+,+)-\sigma(-,-)}{\sigma(+,+)+\sigma(-,-)},\\
\mathcal{A}_2 & = & \frac{\sigma(+,-)-\sigma(-,+)}{\sigma(+,-)+\sigma(-,+)},
\end{eqnarray}
\begin{eqnarray}
\mathcal{A}_3 & = & \frac{\sigma(+,+)-\sigma(-,+)}{\sigma(+,+)+\sigma(-,+)},\\
\mathcal{A}_4 & = & \frac{\sigma(+,-)-\sigma(-,-)}{\sigma(+,-)+\sigma(-,-)},
\end{eqnarray}
\begin{eqnarray}
\mathcal{A}_5 & = & \frac{\sigma(+,+)-\sigma(+,-)}{\sigma(+,+)+\sigma(+,-)},\\
\mathcal{A}_6 & = & \frac{\sigma(-,+)-\sigma(-,-)}{\sigma(-,+)+\sigma(-,-)}.
\end{eqnarray}
\begin{figure*}
\scalebox{0.75}{
\includegraphics{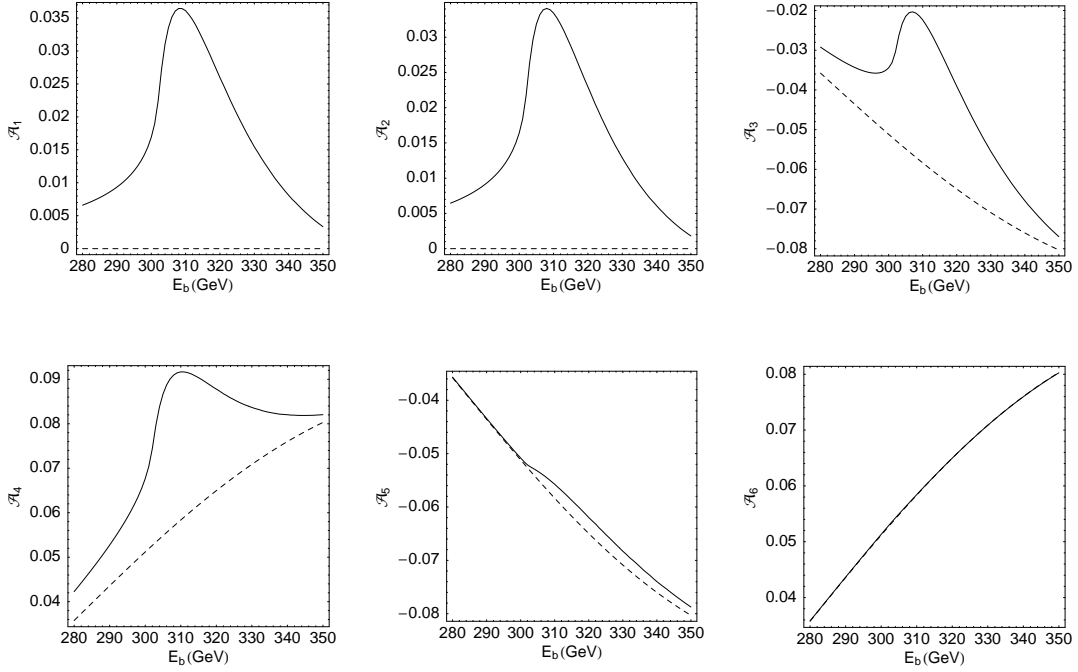}}
\caption{\label{figAsyE}
All six asymmetries are plotted as a function of beam energy $E_b$ for the SM
(dashed line) and the MSSM (solid line) at an angular cut of $\theta_0 = 60^{
\circ}$ in the lab frame. At $E_b$ = 310 GeV, owing to resonance in the 
$s$-channel, the MSSM values of asymmetries are maximally different from that of 
the SM.}
\end{figure*}
Of the above six, $\mathcal{A}_1$ and $\mathcal{A}_2$ are purely CP-violating,
$\mathcal{A}_3$ and $\mathcal{A}_4$ are polarization asymmetries for a given
charge of the lepton, and $\mathcal{A}_5$ and $\mathcal{A}_6$ are charge 
asymmetries  for a given polarization. All these asymmetries are plotted 
against the beam energy
$E_b$ for SM and MSSM in Fig. \ref{figAsyE}. From these plots it is clear 
that $E_b =$ 310  GeV is a good choice for putting limits on the size of the
form factor, for our choice of the  mass of the scalar $m_\phi$.
\subsection{Sensitivity and Limits}
After choosing a suitable beam energy for the analysis, the next thing to look
for is a suitable angular cut in the lab frame, which will  maximise the
sensitivity of the measurement. For asymmetries to be observable, the number of
events corresponding to the asymmetry must be larger than the statistical
fluctuation in the measurement of the total number of events. If $N$ is the 
total number of events then the number of  events corresponding to the asymmetry
must be at least $f\sqrt{N}$, where $f=1.96$ for 95\% C.L. The number of events
$N = \sigma {\cal L}$, where ${\cal L}$ is the luminosity. Asymmetries are
defined as
$${\cal A}= \frac{\sigma_1-\sigma_2}{\sigma_1+\sigma_2}=
 \frac{\Delta\sigma}{\sigma}.$$ 
Thus the number of events corresponding to the asymmetry is ${\cal L }
\Delta \sigma$. For the asymmetry to be measurable at all we must have at least
${\cal L} \Delta \sigma > f \sqrt{{\cal L} \sigma}$, with $f$ denoting the 
degree of significance with which we could  assert the existence of an
asymmetry. Thus the ratio $ ({\cal L} {\Delta \sigma})/
(f \sqrt{{\cal L}\sigma})   = (\sqrt{{\cal L}}/f) \times (\Delta \sigma 
/\sqrt{\sigma})$ will be a measure of the sensitivity. One can be more 
precise in defining this by noting that the fluctuation in the asymmetry is 
given by
$$\delta{\cal A} = \frac{f}{\sqrt{\sigma {\cal L}}}\sqrt{1+{\cal A}^2}\approx
\frac{f}{\sqrt{\sigma {\cal L}}}, $$
for ${\cal A} \ll 1 $. The larger the asymmetry with respect to the 
fluctuations, the larger will be the sensitivity with which it can be measured.
\begin{figure*}
\scalebox{0.75}{
\includegraphics{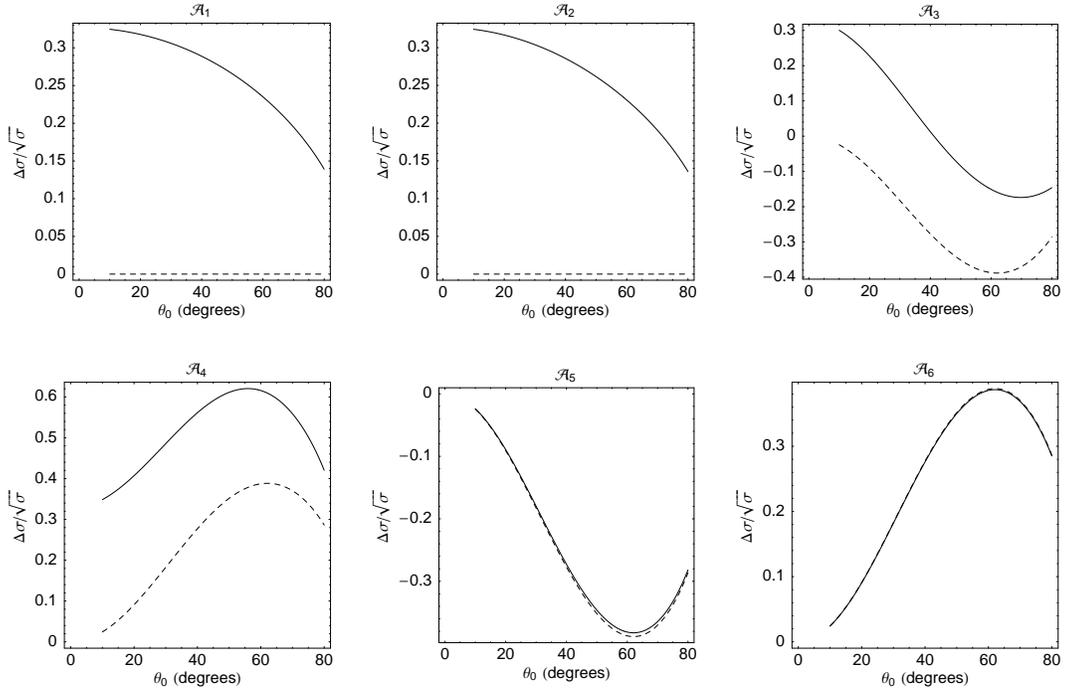} }
\caption{\label{senA}
$\Delta\sigma/\sqrt{\sigma}$, which is proportional to the {\em
sensitivity} $\mathcal{S}$, is plotted against  the angular cut $\theta_0$ for
all six asymmetries. Solid line is for the MSSM and dashed line for the SM.
For $\mathcal{A}_1$ and $\mathcal{A}_2$, the smaller angular 
cut is favourable while for others $\theta_0 = 60^{\circ}$ is a good choice.}
\end{figure*}
We define {\em sensitivity} as,
$${\cal S}=\frac{{\cal A}}{\delta{\cal A}} \propto \frac{\Delta\sigma}
{\sqrt{\sigma}}.$$
$\Delta\sigma/\sqrt{\sigma}$, which is proportional to the sensitivity, 
is plotted for all asymmetries in Fig. \ref{senA} against the angular cut
in the lab frame, $\theta_0$. Since $\mathcal{A}_1$ and $\mathcal{A}_2$ are 
purely CP-violating, they have no contribution from SM for any angular cut. 
Hence for them, the
sensitivity is large when the angular cut is small, because of better statistics.
For the other four there is an SM contribution that varies with the cut.
Though the exact position of the peak in Fig. \ref{senA} depends upon the relative
sizes and signs of the form factors, $\theta_0 = 60^{\circ}$ seems to be a good
choice for the angular cut to maximise the sensitivity of four of the
asymmetries.\\\\
The process under consideration violates CP in general.  But 
when the cut $\theta_0$ is  
$0$, the partial cross sections for $l^+$ and $l^-$ production
become the corresponding total cross sections, and are therefore equal, 
because of charge conservation. Hence for $\theta_0 \rightarrow 0$,
$\mathcal{A}_5$ and $\mathcal{A}_6$ approach zero. In that limit, the
polarization asymmetries $\mathcal{A}_3$ and $\mathcal{A}_4$ are purely 
CP-violating. Thus for $\mathcal{A}_3$ and $\mathcal{A}_4$, apart from the
choice  $\theta_0 = 60^{\circ}$, where the sensitivity peaks,  
$\theta_0 = 0^{\circ}$ would also be a good choice for isolating CP-violating 
parameters.  But, to be away from the beam pipe, we choose the lowest cut to 
be $20^{\circ}$ in the lab frame. We have used
 all six asymmetries for angular cuts of $20^{\circ}$ and $60^{\circ}$ to put 
limits on the combinations $x_i$ and $y_i$. 
\begin{figure*}
\rotatebox{270}{ \scalebox{0.70}{ \includegraphics{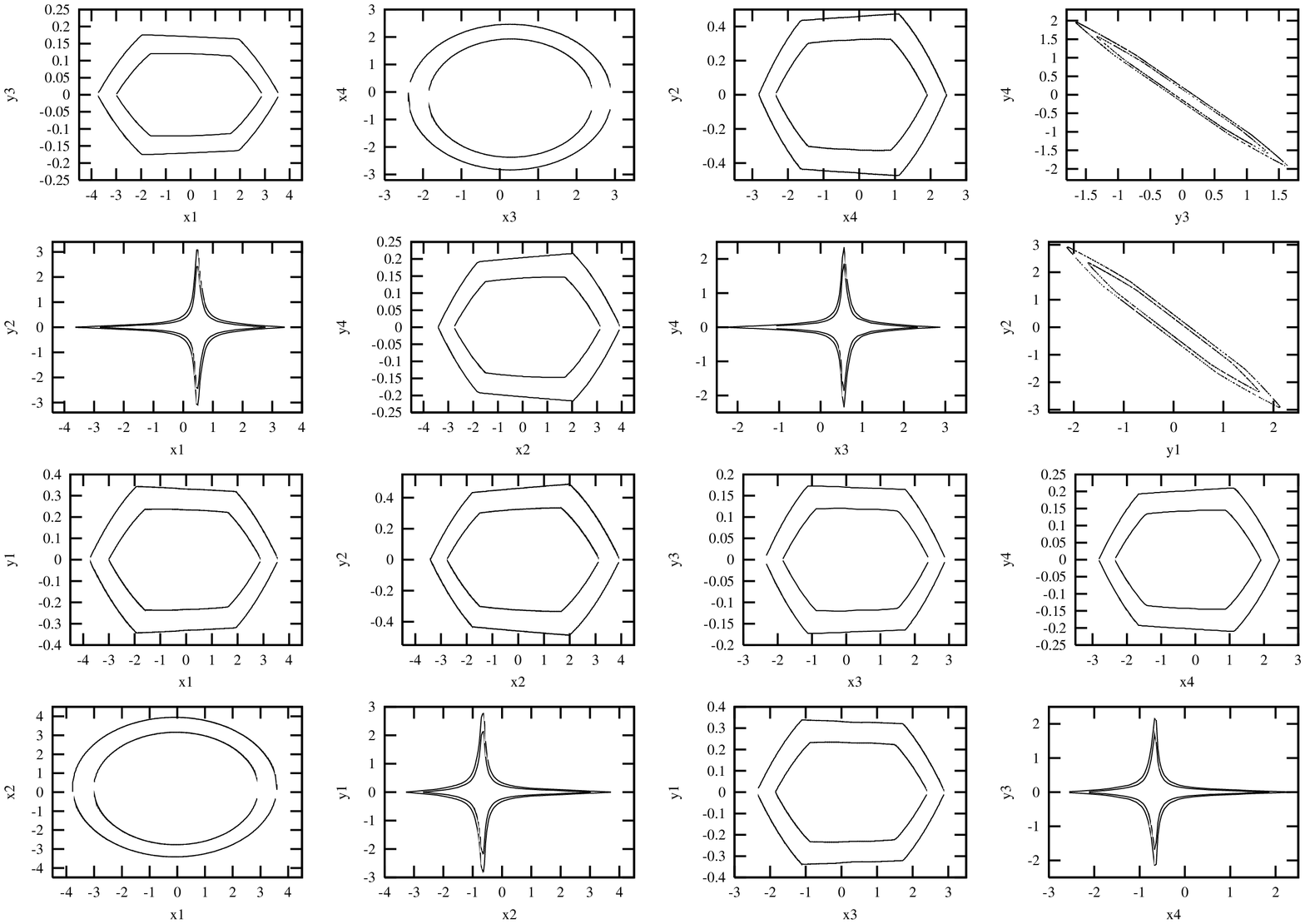} } }
\caption{\label{plim}
The boundaries of blind regions in the parameter space are plotted for
various pairs of parameters, for luminosities 500 and 1000 fb$^{-1}$ 
at beam energy $E_b =$ 310  GeV. Both angular cuts, $\theta_0 = 20^{\circ}$ 
and $60^{\circ}$, are used to put limits at C.L. of 95\%. The larger
region corresponds to 500 fb$^{-1}$, while the smaller corresponds to 
1000 fb$^{-1}$.}
\end{figure*}
\begin{table}
\caption{\label{tab1}List of 95\% C.L. limits on all the
combinations at 500 fb$^{-1}$ and 1000 fb$^{-1}$ at $E_b =$ 310  GeV. These 
limits are obtained from data plotted in Fig. \ref{plim}.}
\begin{ruledtabular}
\begin{tabular}{cccccc}
  &min         &max           &min             &max  &MSSM     \\
  &(500 fb$^{-1}$)&(500 fb$^{-1}$)&(1000 fb$^{-1}$)&(1000 fb$^{-1}$)&value\\
\hline\hline
$x_1$&  $-3.775 $&        3.594 &         $-2.990 $&       2.869 &$-0.429$\\
$x_2$&  $-3.413 $&        3.896 &         $-2.748 $&       3.111 &$-0.396$\\
$x_3$&  $-2.386 $&        2.873 &         $-1.842 $&       2.386 &$-0.077$\\
$x_4$&  $-2.837 $&        2.465 &         $-2.375 $&       1.930 &$+0.165$\\
$y_1$&  $-2.786 $&        2.786 &         $-2.148 $&       2.148 &$-0.168$\\
$y_2$&  $-3.095 $&        3.095 &         $-2.433 $&       2.433 &$+0.363$\\
$y_3$&  $-2.155 $&        2.155 &         $-1.687 $&       1.687 &$-0.195$\\
$y_4$&  $-2.346 $&        2.346 &         $-1.867 $&       1.867 &$-0.180$
\end{tabular}
\end{ruledtabular}
\end{table}
If for certain values of the form factors the asymmetries lie within the
fluctuation from their SM values, then that particular point in the
parameter space cannot be distinguished from SM at a given luminosity. That
point will be said to fall in the blind region of the parameter space. Thus
the set of parameters \{$x_i,y_i$\} will be inside the blind region at a given
luminosity if
\begin{equation}
|{\cal A}(\{x_i,y_i\})-{\cal A}_{\rm SM}| \leq \delta{\cal A}_{\rm SM} =
\frac{f}{\sqrt{\sigma_{\rm SM} {\cal L}}}\sqrt{1+{\cal A}_{\rm SM}^2}.
\label{blindeqn}
\end{equation}
For simplicity we have taken only two of the eight combinations 
to be non-zero at a time and
have constrained them in 16 different planes, shown in Fig. \ref{plim},
satisfying their inter-relations. The limits obtained on each of the 
combinations by taking a union of the blind regions in the 16  plots are
listed in Table \ref{tab1}. Also 
shown in the last column of the table are the values of $x_i,y_j$  for 
the MSSM point we have chosen for Figs. \ref{figAsyE} and \ref{senA}.\\\\
Next we do a small exercise to see whether these asymmetries have the potential 
to distinguish between the SM and MSSM. It is clear that we can repeat the
analysis of finding blind regions in the ($x_i,y_j$)  planes around a
particular point predicted by the MSSM.  The values of $x_i,y_j$,
corresponding to our choice of the MSSM point given by Eq. (\ref{mssmparam})  
are listed in last column of Table \ref{tab1}.
The blind regions around these values will be defined by an equation 
similar to Eq. (\ref{blindeqn}), where ${\cal A}_{\rm SM}$ will be replaced
by ${\cal A}_{\rm MSSM}$. In ${\cal A}(x_k,y_l)$ all independent $x_k,y_l$  
other than the pair 
$x_i,y_j$  being considered, are set to their MSSM value, and the pair 
$x_i,y_j$ is then varied. We show in 
\begin{figure}
\rotatebox{0}{ \scalebox{0.70}{ \includegraphics{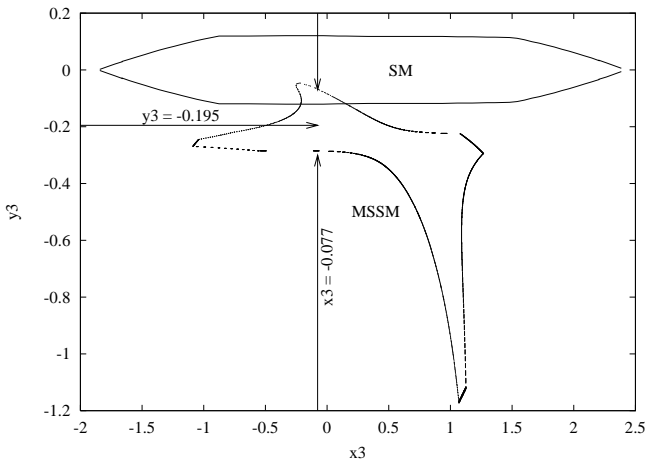} } }
\caption{\label{smmssm}
The boundaries of  blind regions in the parameter space are plotted in
$x_3-y_3$ plane, for a luminosity of 1000 fb$^{-1}$ at beam energy
$E_b =$ 310 GeV. Both angular cuts, $\theta_0 = 20^{\circ}$ and $60^{\circ}$,
are used to put limits at C.L. of 95\%. 
For this MSSM point, $x_3=-0.077, \ y_3=-0.195$.}
\end{figure}
Fig. \ref{smmssm} the results of such an exercise for the parameter pair
($x_3,y_3$) along with the corresponding one for the SM. This  shows that
these studies can be sensitive to the CP  mixing produced by loop effects.
Of course one needs to study this over the supersymmetric parameter space.
But the example shown here clearly shows the promise of the method.
%
%
\section{Discussion}
\label{six}
The four cross sections depending on the polarization of the initial lepton 
and the charge of the final state lepton that we  use to construct 
asymmetries, can in general be written as
\begin{equation}
\sigma(\lambda_e,Q_l) = \sigma_{00} + \sigma_{01} Q_l + \sigma_{10} \lambda_e
+ \sigma_{11} Q_l \lambda_e.
\label{4sig}
\end{equation}
This says that we have four independent $\sigma_{ij}$, which constitute four
polarized cross-sections. Out of these four, $\sigma_{00}$ is the largest and
others are of the order of a few per cent of $\sigma_{00}$. Thus we can safely
approximate denominators of all ${\cal A}_i$'s to $\sigma=2\sigma_{00}$. 
This makes ${\cal A}_i$'s proportional to their numerators, which consists of 
only three of $\sigma_{ij}$.
Thus out of six asymmetries constructed in Section \ref{five} only three are 
independent and we cannot determine all six form factors simultaneously using 
these asymmetries. This is a reflection of the fact that there are only 
three CP-violating asymmetries~\cite{wengan} at the production level of the 
$t \bar t$ pair; one is for the unpolarized case, and the other two are 
polarization 
asymmetries. The ${\cal A}_i$'s defined here are combinations of these 
three.\\\\
In Fig.~\ref{plim} we took only two combinations as non-zero and varied them to
find the blind region in that plane. We found strong limits on $y$'s and almost
no limits on $x$'s in each of the planes. When we allowed three of the
combinations to vary simultaneously there were almost no limits on any of the
combinations. This can be understood by looking at
Eq. (\ref{4sig}). The charge asymmetries are very small and approach zero as
we reduce the angular cut, which implies that $\sigma_{01}$ and $\sigma_{11}$
are very small and tend to zero as $\theta_0\rightarrow 0$. Thus two of the
four independent components of the polarized cross-section are very small 
(typically by a factor of 100 to 500); neglecting them, we are left with only 
two independent components, implying that only one of the six asymmetries is 
independent.
Thus, though we have four independent components at hand, two being small we 
are effectively left with only two almost identical strong constraints, and
thus essentially only one. These asymmetries thus  constrain only $y$'s and 
leave $x$'s mostly unconstrained. The fact that $x$'s are constrained at all is
because the equation of the boundary of a blind region arising from any
one asymmetry, for two variables, is an
equation of a pair of conic sections. The blind regions shown in Fig. \ref{plim}
are intersections of blind regions obtained from all six asymmetries with two 
different angular cuts.
\subsection{The Strategy}
All the cross sections and asymmetries are expressible in terms of $x$'s 
and $y$'s alone and, of these, only five are independent.  Thus any number of 
asymmetries for any general polarization can never determine
all six form factors as only five independent combinations appear in the
expressions. For $S_t$ and $P_t$ we have to rely on partial decay width 
measurements of the scalar $\phi$ to $t\bar t$ pair. 
Thus, if we have a few more independent and strong constraints,
we will be able to put simultaneous limits on all six form factors.
But with circular polarization we have only the four observables used here 
already. 
One possibility would be to use the dependence of the angular distribution of 
the decay leptons on the initial state photon polarization. But to do that one 
would  need a large statistics, which will not be available even with
an integrated luminosity of $10^3$ fb$^{-1}$. The other option is to look for 
linearly polarized initial photons. Here, by choosing different angles between 
the planes of polarizations one  can alter the relative contribution from 
CP-even and CP-odd Higgs. This, along with the asymmetries considered and the 
partial decay width of $\phi$, can then be used to put limits on all six form 
factors simultaneously or alternatively  to determine them. Some discussions 
of these for the $t \bar t$ production exist already~\cite{asakawa}
\\\\
In view of above analysis we can propose a strategy for characterizing the
heavy scalar $\phi$. The first step would be to determine its mass $m_{\phi}$,
its total decay width $\Gamma_{\phi}$ and the partial decay width to a 
$t\bar t$ pair.
The last will tell us about $S_t^2+P_t^2$. Then the second step will be to look
for asymmetries ${\cal A}_1$ and ${\cal A}_2$ to see if there is any CP
violation. Step 3 depends upon the outcome of step 2. In case of 
non-observation of CP violation, one will have to look for linearly polarized
asymmetries to see whether the Higgs is CP-even or CP-odd. If CP violation
is observed, then all the asymmetries, for circular and linear polarizations, 
can be used to determine the form factors.
\subsection{Discriminating Models}
As we have seen, it is not possible to determine all the combinations 
$x_i,y_j$ using the asymmetries we have constructed. However, as discussed
below, we can surely use the model predictions for these to discriminate
against a particular model when data are available or test the possibilities
of being able to distinguish between different models at a given luminosity.
\\\\
The blind region around any model point in the five-dimensional parameter space 
is a non-convex 
structure and extends far out from the model point in some of the directions. 
Thus projection on any plane may result in a large blind region, which can be 
misleading. Thus it is not possible to restrict to less than the full set
of 5 parameters for testing models. Below we develop a method for
distinguishing between models and checking whether they are ruled out by
experiment.\\\\
The simplest way to compare two models is to ask if the first model
point lies within the blind region of the second and vice versa. If not, we say
that the model predictions are distinguishable from each other at the 
luminosity and confidence level considered.
As an example, we chose two models - SM and 
MSSM. The MSSM is same as given by the last column of
Table \ref{tab1}. 
\begin{figure}
\scalebox{0.75}{
\includegraphics{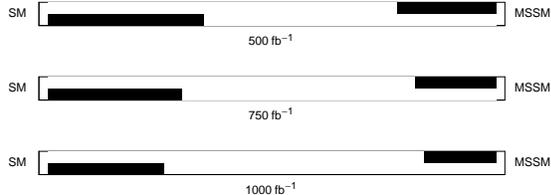}}
\caption{\label{models}
Blind regions along the line joining two model points, SM and MSSM, in the 
five-dimensional parameter space at ${\cal L}$ = 500, 750 and 1000 fb$^{-1}$.
All six asymmetries with both angular cuts, $\theta_0=20^0$ and $60^0$, are
used.}
\end{figure}
The model points in the five-dimensional
parameter space are connected by a line parameterized by $t$ with
$t=0$ corresponding to one model and $t=1$ to the other.  
We have calculated the blind region
around each of the models along the connecting line. These are shown 
in Fig. \ref{models} and it can be seen that
each model sits well outside the blind region of the other at an
integrated luminosity of 500 fb$^{-1}$. Furthermore,
their blind regions do not overlap along these lines. Thus we can say that
our method can distinguish candidate models at a certain luminosity (500
fb$^{-1}$ in this case).
A more accurate way will be to search the whole of the five-dimensional 
parameter space for the overlap
of the blind regions corresponding to two candidate models and not just 
along the line joining them. If no such overlap is 
found then we can say for sure that the models can be distinguished. This search
could be quite complicated. Alternatively we can use the numerical values of 
individual asymmetries and fluctuations directly as discussed below.\\\\
It is clear that we can determine blind region around a given model prediction
in any parameter space given the numerical value of the model predictions for
asymmetries and the statistical fluctuations expected in it at a given
luminosity. Any change in these numerical values will yield a different blind 
region in the five-dimensional parameter space. One will then test asymmetry 
predictions for a particular model against an experimental measurement or 
compare the predictions of two models against each other to draw conclusions 
about their distinguishability at a given luminosity and confidence level. 
%
\begin{table}
\caption{\label{tab2}
The probability ${\cal P}$ of confusing SM with MSSM at 95\% 
C.L. for different asymmetries and luminosities. 
}
\begin{ruledtabular}
\begin{tabular}{cccc}
Asymmetries	& ${\cal P}$ at	500 fb$^{-1}$ &{\cal P} at 750 fb$^{-1}$ &${\cal P}$ at 1000 fb$^{-1}$\\ 
\hline\hline
${\cal A}_1(\theta_0=60^0)$ &$5.5\times10^{-4}$ &$4.8\times10^{-6}$	& $6\times 10^{-8}$ \\
${\cal A}_2(\theta_0=60^0)$ &$7.8\times10^{-4}$ &$8.0\times10^{-6}$	& $7\times 10^{-8}$ \\
${\cal A}_3(\theta_0=60^0)$ &$3.4\times10^{-4}$ &$2.0\times10^{-6}$ 	& $1\times10^{-8}$ \\
${\cal A}_4(\theta_0=60^0)$ &$1.4\times10^{-3}$ &$2.1\times10^{-6}$	& $2.8\times10^{-7}$ \\
${\cal A}_1(\theta_0=20^0)$ &$1.6\times10^{-7}$ &$<10^{-8}$	& $<10^{-8}$ \\
${\cal A}_2(\theta_0=20^0)$ &$2.3\times10^{-7}$ &$<10^{-8}$	& $<10^{-8}$ \\
${\cal A}_3(\theta_0=20^0)$ &$2.3\times10^{-7}$ &$<10^{-8}$	& $<10^{-8}$ \\
${\cal A}_4(\theta_0=20^0)$ &$2.2\times10^{-7}$ &$<10^{-8}$	& $<10^{-8}$ \\
\end{tabular}
\end{ruledtabular}
\end{table}
If the values of asymmetries expected at the particular level of confidence,
corresponding to (say) two models, have no overlap, then the two models are
distinguishable at that confidence level.
There is
still a non-zero probability that the models can be confused with each other 
in an experiment. To determine the probability of such a confusion we
take any one asymmetry at a time and calculate the limits upto which the
predicted asymmetry values can fluctuate at a certain level of confidence 
in the models under consideration.
Then we generate normally distributed random numbers centered at the
asymmetry corresponding to the first model, say SM, with standard deviation
same as the 1 $\sigma$ fluctuation of the SM asymmetry. We count the number of
points for which the asymmetry value lies within the 95\% confidence interval of the other model, say
MSSM. The number of such points divided by the total number of points taken
is the probability ${\cal P}$ of confusing SM with MSSM at 95\%
confidence level. As Table \ref{tab2} indicates, ${\cal P}$ is of the order
of $10^{-7}$ for a luminosity of 500 fb$^{-1}$, and we can safely say that SM is
distinguishable from MSSM at 500 fb$^{-1}$. In a similar way we can replace
SM by the experimental asymmetries and MSSM by a candidate model. Now
even if for one of the asymmetries ${\cal P}$ is very small ($O(10^{-3})$) 
we can simply
reject the model as in words it translates to: {\em the probability of the 
experimental results being statistical fluctuation of the candidate model
at 95\% C.L. is very small.} In fact, the method described above is nothing
but the {\em Step 2} of our strategy discussed in previous sections, when one
talks about asymmetries ${\cal A}_1$ and ${\cal A}_2$. 
%
%
\section{Conclusion}
\label{seven}
We have demonstrated how the Higgs-mediated CP violation in the process  
$\gamma \gamma \rightarrow t\bar t$ can be studied by looking at the 
integrated cross section of $l^+/l^-$ coming from the decay of $t/\bar t$. 
We demonstrated that the decay lepton angular distribution is insensitive 
to any anomalous part of the $tbW$ coupling, $f^{\pm}$, to first order. 
We constructed  combined asymmetries involving  the initial lepton (and hence
the photon) polarization and the decay lepton charge. We showed that using
only circularly polarized photons will be inadequate to determine or 
constrain the sizes of all form factors simultaneously, but can put strong 
limits on CP-violating combinations, $y$'s, when only two combinations are 
varied at a time. 
We show, by taking an example of a particular choice of MSSM parameters, 
that the analysis is sensitive to the CP mixing at a level that is 
generated by loop effects.  
We also further sketch a possible strategy to characterize the scalar 
$\phi$ using linear polarization.  \\\vspace{0.2cm}\\
\centerline{ACKNOWLEDGEMENT}\\\\
We thank Prof. N. V. Joshi for useful discussions.
\appendix
\section{Expressions for $A_{ij}$, $B_{ij}$, etc.}
For circularly polarized photons the form factors $A_{ij}$ and $B_{ij}$ are
given below; $\eta_2$ and $\xi_2$ are the degrees of circular polarization of
two colliding photons:
\begin{eqnarray*}
A_{00}^{\pm}&=&2|A_{\phi}|^2(\beta^2S_t^2+P_t^2)\left[ (|S_{\gamma}|^2 +
|P_{\gamma}|^2)\right.\nonumber\\
&&\left.\left(\frac{1+\eta_2\xi_2}{2}\right) + 2\Im(S_{\gamma}P^*_
{\gamma})\left(\frac{\eta_2 + \xi_2}{2}\right)\right]\\
A_{01}^{\pm}&=&4A_c\left[(\beta^2S_t\Re(A_{\phi}S_{\gamma}) + P_t
\Re(A_{\phi}P_{\gamma}))\left(\frac{1+\eta_2\xi_2}{2}\right) \right.\nonumber\\
&+& \left.(P_t
\Im(A_{\phi}S_{\gamma})-\beta^2S_t\Im(A_{\phi}P_{\gamma})) \left(\frac{\eta_2 +
\xi_2}{2}\right)\right]\\
A_{02}^{\pm}&=&2A_c^2\left[ (1+\beta^2)\left(\frac{1+\eta_2\xi_2}{2}\right)
\right.\nonumber\\
&+&\left.\frac{\beta^2(2-\beta^2)}{1-\beta^2}\left(\frac{1-\eta_2\xi_
2}{2}\right)\right]\\
A_{22}^{\pm}&=&-4A_c^2\beta^2\left(\frac{1-\eta_2\xi_2}{2}\right)\\
A_{42}^{\pm}&=&-\frac{2A_c^2\beta^4}{1-\beta^2}\left(\frac{1-\eta_2\xi_
2}{2}\right)
\end{eqnarray*}
\begin{eqnarray*}
B_{01}^{\pm}&=&4\beta A_c\left[ (S_t\Re(A_{\phi}S_{\gamma}) + P_t
\Re(A_{\phi}P_{\gamma}))\left(\frac{\eta_2 + \xi_2}{2}\right)
\right.\nonumber\\ 
&+& \left.(P_t\Im(A_{\phi}S_{\gamma}) - S_t\Im(A_{\phi}P_{\gamma}))
\left(\frac{1+\eta_2\xi_2}{2}\right)\right]\\
B_{02}^{\pm}&=&4\beta A_c^2\left(\frac{\eta_2 + \xi_2}{2}\right)\\
B_{12}^{\pm}&=&\frac{4A_c^2\beta^2}{1-\beta^2}\left(\frac{\eta_2 -
\xi_2}{2}\right)= \ B_{12}\\
B_{32}^{\pm}&=&-B_{12}^{\pm}=-B_{12}
\end{eqnarray*}
\begin{eqnarray*}
C_0 &=&  4\beta^2A_c^2 \left(\frac{\eta_2 -\xi_2}{2}\right)\\
D^{\pm}&=&0\\
A_c&=&\frac{4Q_t^2m_t}{\sqrt{s}} = 2Q_t^2\sqrt{1-\beta^2}\\
A_{\phi}&=&\frac{e}{16\pi^2} \ \frac{m_t}{m_W} \ \frac{s}{s-m_{\phi}^2+
im_{\phi}\Gamma_{\phi}}
\end{eqnarray*}
\begin{eqnarray*}
X_0&=&2+\beta^2\sin^2\theta_l\\
X_1&=&-4\beta\cos\theta_l \\
X_2&=&\beta^2(3\cos^2\theta_l-1)\\
Y_0&=&-\cos\theta_l(2+\beta^2\sin^2\theta_l)\\
Y_1&=&\beta(3+\cos^2\theta_l)\\
Y_2&=&2\cos\theta_l(2-\beta^2\cos^2\theta_l)\\
Y_3&=&-2\beta(3+\cos^2\theta_l)\\
Y_4&=&\cos\theta_l(-2+3\beta^2+\beta^2\cos^2\theta_l)\\
Y_5&=&\beta(3+\cos^2\theta_l)\\
Y_6&=&-2\beta^2\cos\theta_l
\end{eqnarray*} 

\end{document}